\documentclass[%
 reprint,
prx,
]{revtex4-2}
\usepackage{graphicx}
\usepackage{xcolor}
\usepackage{dcolumn}
\usepackage{bm}
\usepackage{amsmath, amssymb}
\usepackage{diagbox}

\usepackage{hyperref}
\usepackage{url}
\usepackage[utf8]{inputenc}
\begin{document}

\preprint{APS}
\title{Landau-Level-Resolved Mode Mixing and Shot Noise in Gate-Defined Graphene Quantum Point Contacts}
\author{Shakthidhar Vilvanathan}
 \altaffiliation{Current address: Indian Institute of Science Education and Research Thiruvanathapuram, Thiruvananthapuram, Kerala, India}
 \affiliation{Low Temperature Laboratory, Department of Physics, Aalto University, Espoo, Finland.}
 
\author{Jerin Saji}%
\affiliation{Low Temperature Laboratory, Department of Physics, Aalto University, Espoo, Finland.}
\author{Kristiana Frei}%
\affiliation{Low Temperature Laboratory, Department of Physics, Aalto University, Espoo, Finland.}

\author{Jakub Tworzydło}%
\email{jakub.tworzydlo@fuw.edu.pl}
\affiliation{Faculty of Physics, University of Warsaw, Warsaw, Poland.}
\author{Manohar Kumar}
\email{manohar.kumar@aalto.fi}
\affiliation{Low Temperature Laboratory, Department of Physics, Aalto University, Espoo, Finland.}
\affiliation{QTF Centre of Excellence, Department of Applied Physics, Aalto University, P.O.
Box 15100, FI-00076 Aalto, Finland.}

\begin{abstract}
Graphene quantum point contacts (QPCs) in the quantum Hall regime host several competing transport mechanisms, including chiral edge propagation, valley degeneracy, and gate-induced mode mixing. Their interplay is not visible in conductance alone. Shot noise directly probes the statistics of transmission eigenvalues and thereby reveals microscopic mode partitioning, which conductance measurements cannot access. We develop a hybrid theoretical framework combining realistic tight-binding simulations of gate-defined graphene QPCs with random matrix theory (RMT) to predict shot noise and Fano factor signatures across different transport regimes. We validate the tight-binding model against experimental conductance maps of hBN-encapsulated graphene Hall bars. The model reproduces mode-selective transmission and identifies three distinct regimes: adiabatic propagation, sharp mode filtering, and multi-mode mixing driven by localized states beneath the split gate. Translating these microscopic transmission eigenvalue distributions into noise statistics via RMT, we find that the Fano factor exhibits a universal crossover as a function of the Landau level index. For higher Landau levels ($N_L > 0$), complete mode mixing in an effectively chaotic cavity produces an asymptotic saturation limit of $F \simeq 1/4$, consistent with RMT predictions for a chaotic scatterer. In contrast, the zeroth Landau level ($N_L = 0$) converges to a limit of $F = 1/3$. We show that this distinct value has a microscopic origin in the sublattice polarization of the $N_L = 0$ edge state. The generically mixed-sublattice localized states beneath 
the gate are strongly suppressed, confining transport to an effective single channel ($N = 1$). Complete mixing within this single channel yields a flat eigenvalue distribution and hence exactly $F = 1/3$, the same numerical value as pseudo-diffusive zero-field graphene transport, but arising here from single-channel RMT rather than Dirac-point evanescent mode physics. The $F = 1/3$ versus $F=1/4$ crossover constitutes a Landau-level-resolved noise signature that is absent in conductance. It provides a direct discriminator between multi-channel chaotic mixing and single-channel transport in graphene QPCs, and offers quantitative benchmarks for experimental noise measurements.

\end{abstract}

\keywords{Quantum Hall transport, Random matrix theory, shot noise, quantum point contact, and graphene}

\maketitle


\section{\label{sec:int}INTRODUCTION}
Graphene provides an unusually rich setting for studying quantum transport. Its massless Dirac spectrum, ambipolar gating, and high carrier mobility make it possible to reach regimes that conventional two-dimensional systems cannot easily access~\cite{Novoselov2005, Neto2009}. When a large magnetic field is applied, graphene develops relativistic Landau levels whose sequence reflects both spin and valley degrees of freedom~\cite{Schakel1991}. For a Landau level 
$N_L$, the quantum Hall plateaus follow the characteristic relativistic sequence $\sigma_{xy} = 4e^2/h(N_L+1/2)$, where the factor of four reflects spin and valley degeneracy~\cite{Gusynin2005}. Each Landau level gives rise to a corresponding quantum Hall edge channel that follows the physical boundaries of the graphene two-dimensional electron gas (2DEG).

\par With electrostatic gating techniques, it is now possible to locally 
structure the potential landscape at the scale of a few tens of nanometers, allowing unprecedented control over how individual quantum Hall edge channels form, propagate, and interact~\cite{Williams2007, Ozyilmaz2007, Ki2009, Nakaharai2011, Xiang2016, Zimmermann2017, Ahmad2019, Cohen2023, Pandey2024}. Among the most remarkable developments is the realization of graphene p--n 
junctions, where electron-like and hole-like edge states meet along a gate-defined interface, acting as tunable mixing regions for chiral modes and producing equilibration behaviors with no analogue in conventional Hall systems~\cite{Ki2009, Amet2014}. Building on this, split-gate quantum point contacts (QPCs) exploit the energy gaps between Landau levels to enable precise tuning of individual edge channels~\cite{Zimmermann2017, Cohen2023, Pandey2024}. At the common boundary, chiral modes propagating in the same or opposite directions exchange carriers and relax to a shared electrochemical potential, giving rise to partial, full, or selective equilibration and conductance plateaus at half-integer filling factors that can mimic signatures of non-Abelian states~\cite{Ki2009b, Pandey2024}.

\par Shot noise has played a central role in revealing this physics. Unlike conductance, which reflects only the average transmission, shot noise directly probes fluctuations and the microscopic structure of current-partitioning~\cite{Blanter2000}. Early studies showed that even without a magnetic field, noise measurements expose evanescent channels and pseudo-diffusive transport near the Dirac point~\cite{Danneau2008, Williams2007}. In the quantum Hall regime, p--n junctions behave as tunable electronic beam splitters, capable of generating finite noise even when conductance is perfectly quantized~\cite{Abanin2007}, with the Fano factor sensitive to junction length, disorder, and equilibration between co-propagating modes~\cite{Kumada2015, Matsuo2015}. Elastic mixing tends to produce universal noise values~\cite{Abanin2007}, while inelastic relaxation rapidly suppresses them~\cite{Nagaev1992}.

\begin{figure}
\includegraphics[width=0.4\textwidth,keepaspectratio]{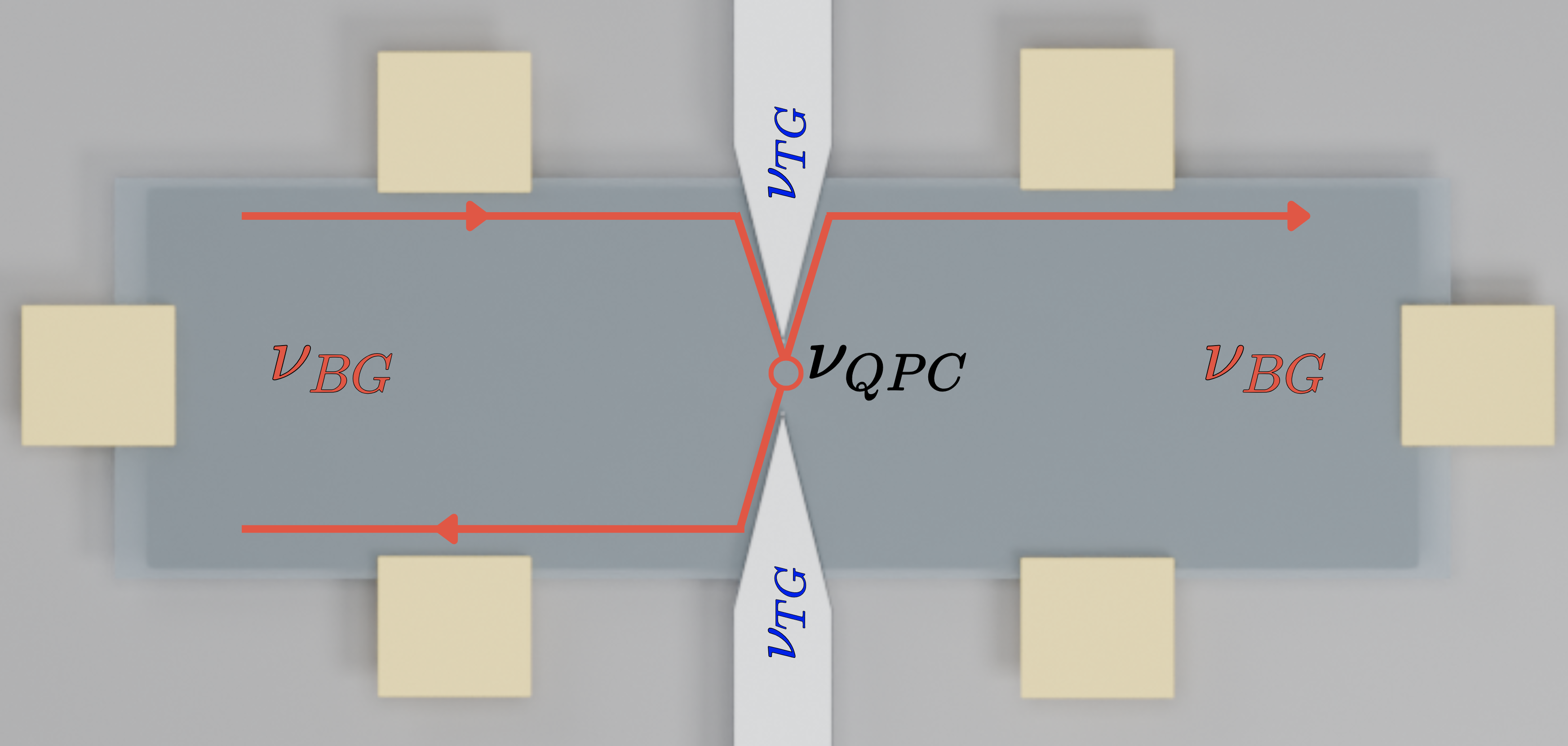}
\caption{\textbf{Schematic representation of the Hall bar sample with a QPC and magnetotransport characterization:}. The different filling factors are segregated by the back gate potential $\nu_\mathrm{bg}$, top gate potential $\nu_\mathrm{tg}$, and local potential within the confined region of QPC $\nu_{QPC}$. The edge state propagating through the QPC is shown and marked with the arrow.\label{fig1}}
\end{figure}

\par This competition becomes even more pronounced in QPCs, where a narrow constriction selectively transmits a subset of incoming modes. Unlike traditional GaAs/AlGaAs heterostructure QPCs, localized or transmitting edge modes persist under the local top gates, and QPC-defined graphene constrictions can support edge states with mixed valley character, exhibit soft electrostatic profiles, and host reconstructed or compressible-edge regions~\cite{Cohen2023, Pandey2024}. These microscopic features can enhance, suppress, or qualitatively reshape shot noise in ways not captured by simple single-particle pictures. Most recently, Garg \textit{et al}.~\cite{Garg2025} reported enhanced shot noise in graphene QPCs with electrostatic reconstruction, attributing the effect to correlated charge tunneling. However, the Landau-level dependence of the noise statistics and, in particular, whether the zeroth and higher Landau levels realize distinct transport universality classes within the same device was not addressed.

\par This open question demands a framework that goes beyond average transmission and captures the full statistical distribution of transmission eigenvalues across different Landau level regimes. Random-matrix theory (RMT) naturally provides such a framework: it predicts universal eigenvalue distributions, connects them to measurable quantities such as shot noise and higher current cumulants~\cite{Beenakker1997}, and classifies transport into distinct universality classes whose noise signatures are robust against microscopic details~\cite{Baranger1994}. Yet applying RMT to graphene QPCs in the quantum Hall regime is not straightforward: graphene combines chiral transport with valley and spin degeneracy, while a QPC introduces finite dwell times, partial and full mode mixing, and sensitivity to long-range Coulomb interactions. Most critically, the sublattice structure of graphene's Landau levels, which differs fundamentally between $N_L = 0$ and $N_L > 0$ introduces a degree of freedom with no analogue in conventional semiconductor QPCs, whose consequences for noise universality classes have not been systematically explored.

\par These considerations raise three interconnected questions that motivate this work. First, what is the microscopic origin of mode mixing in gate-defined graphene QPCs --- where exactly does mixing occur, and how does current partition between transmitted and reflected channels evolve across different filling factor regimes? Second, how does equilibration between co-propagating edge channels depend on the interplay between bulk and QPC filling factors, and what role do localized states beneath the split gate play? Third, and most directly: does the zeroth Landau level, with its unique sublattice-polarized character, realize a fundamentally different universality class of noise statistics than higher Landau levels within the same device, and can this distinction be made quantitative and experimentally accessible via the Fano factor?

\par To address these questions, we employ a hybrid theoretical approach combining microscopic modeling with statistical universality. Tight-binding simulations provide device-realistic transmission eigenvalue spectra, compute the shot noise and Fano factor directly via the Landauer--B\"uttiker formalism, and validate the model against experimental conductance maps. Random-matrix theory then provides the analytical framework that interprets these numerical results, identifies the universality class underlying each transport regime, and demonstrates that the computed Fano factor values reflect deep statistical properties of the transmission eigenvalue distributions rather than accidental features of the device geometry. We demonstrate that a single gate-defined graphene QPC hosts two distinct transport universality classes simultaneously 
accessible by tuning the Landau level index: the zeroth Landau level realizes $F \simeq 1/3$ from single-channel chaotic mixing arising from its sublattice-polarized character, while higher Landau levels realize $F \simeq 1/4$ from many-channel chaotic cavity statistics. This Landau-level-resolved crossover between 
universality classes, absent in both conventional semiconductor QPCs and zero-field graphene, constitutes a uniquely relativistic noise signature with no analogue in non-Dirac two-dimensional systems.

\section{\label{sec:level1}Experimental Platform and Mode-Resolved Transport}

\begin{figure*}
\includegraphics[width=\textwidth,height=0.8\textheight,keepaspectratio]{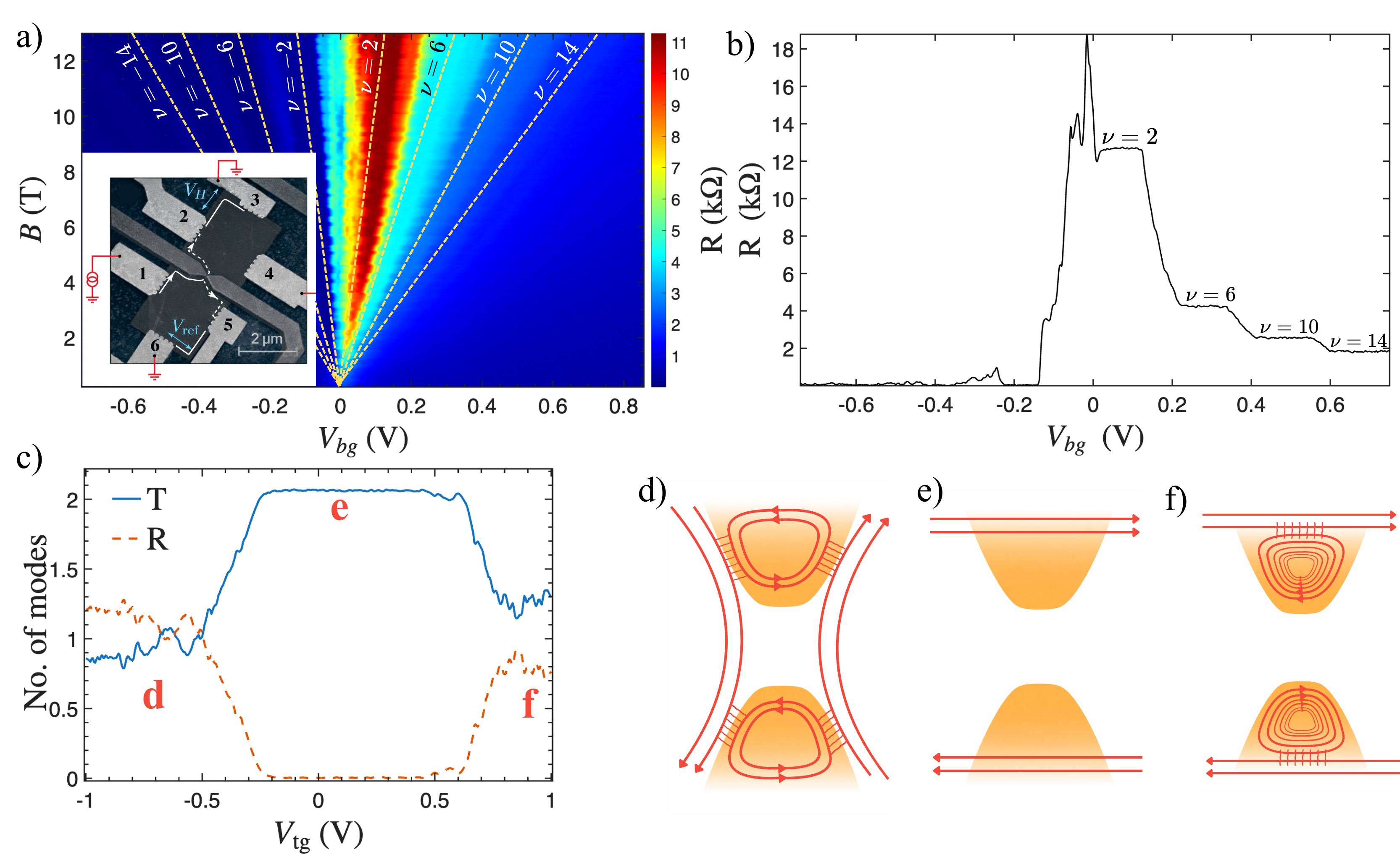}
\caption{\textbf{The experimental results obtained from a characterization of the sample} a) Landau fan diagram: The Hall resistance plotted with respect to the backgate potential and magnetic field. Inset fig shows the representative sample. b) The transmission result with varying gate voltage showing the quantum Hall plateaus of graphene c) The transmission and reflection measurement across the QPC for $\nu_\mathrm{bg} = 2$ filling factor. The transmission resistance is measured across contacts 2 and 3, while reflected resistance is measured at contacts 5 and 6.   d), e), f) representations of how the edge states propagate at different top gate voltages across the $\nu_\mathrm{bg} = 2$ filling factor region.\label{fig2}}
\end{figure*}

Hall-bar devices with dimensions of $10~\mu\mathrm{m}\times5~\mu\mathrm{m}$ and six ohmic contacts were fabricated from hexagonal-boron-nitride–encapsulated graphene heterostructures with graphite back gates (Fig.~\ref{fig2}a). A symmetric metallic split top gate of width $100$~nm was positioned between the two inner voltage probes to define a quantum point contact (QPC). Two nominally identical Hall-bar samples were measured to confirm reproducibility; in the following, we present data from a single representative device. Details of the device fabrication and sample classification are provided in the Methods section.

As shown in Fig.~\ref{fig1}, the Hall bar hosts spatially separated filling factor regions defined by electrostatic gating. The bulk filling factor is controlled by the back gate voltage \(V_{BG}\), while the QPC region is additionally tuned by the top gate voltage \(V_{TG}\). Consequently, the local filling factor at the constriction (\(\nu_{QPC}\)) is determined by the combined action of both gates.

The Hall-bar device was characterized using a low-frequency differential lock-in technique. A current of $1$~nA was injected through contact~1, and the transmitted voltage $V_T$ was measured between contacts~2 and~3, yielding $V_T = I_T R_Q/\nu_{bg}$. The reflected voltage $V_R$ was measured between contacts~5 and~6 and is given by $V_R = I_R R_Q/\nu_{bg}$ (see the inset of Fig.~\ref{fig2}a). Here, $I_T$ and $I_R$ denote the transmitted and reflected currents, respectively, $R_Q$ is the quantum resistance, and $\nu_\mathrm{bg}$ is the back-gate filling fraction. The Landau fan diagram extracted from the transmitted signal is shown in Fig.~\ref{fig2}a. Well-developed quantum Hall plateaus are observed, following the sequence $\nu = 4(N_L + 1/2)$, where $N_L$ is the Landau-level index. The factor of $4$ reflects the combined spin and valley degeneracy of the graphene two-dimensional electron gas (2DEG), indicating that the degeneracy of the edge states remains intact.

The trajectories of the electronic edge channels are schematically illustrated in the inset of Fig.~\ref{fig2}a. For hole-like carriers, \textit{i.e.}, at positive gate voltages, the direction of edge-channel propagation is reversed relative to that shown. Since contact~6 is grounded, signals reaching this contact are fully shunted. Consequently, the quantum Hall plateau signature is suppressed, and only a weak modulation of the resistance resembling the longitudinal resistance $R_{xx}$ is observed depending on the occupation of Landau levels (Fig.~\ref{fig2}b).

We further characterize the QPC by measuring the transmitted $N$ and reflected modes current $M-N$. Here, the filling factor defines $\nu_\mathrm{bg} = M$. The Fig.~\ref{fig2}c shows the transmitted and reflected trace. To extract the number of transmitted and reflected quantum Hall edge modes from the multi-terminal transport measurements, we employ the Landauer--Büttiker formalism for chiral edge transport. 

From the measured resistances $R_{32}$ and $R_{65}$, the number of transmitted and reflected modes can be directly extracted as

\begin{equation}
N = \frac{h}{e^2}\,
\frac{R_{32}}{(R_{32}+R_{65})^2},
\end{equation}
\begin{equation}
M - N = \frac{h}{e^2}\,
\frac{R_{65}}{(R_{32}+R_{65})^2}.
\end{equation}

This procedure enables a direct quantitative extraction of the numbers of transmitted and reflected edge channels from multi-terminal voltage measurements, forming the basis of the conductance maps.

Figure~\ref{fig2}c demonstrates that edge-mode transmission depends on both the bulk filling fraction $\nu_\mathrm{bg}$ and the local filling fraction $\nu_\mathrm{tg}$ set by the back and top gates. For $\nu_\mathrm{bg}=\nu_\mathrm{tg}=2$, the chiral edge state is fully transmitted through the QPC. Tuning the top-gate voltage toward $\nu_\mathrm{tg}=-2$ ($V_{\mathrm{tg}}<0.1$~V) forces the incoming edge state to follow the equipotential contours of the confined region, resulting in increased reflection. Even at the most negative gate voltages, complete reflection is not observed; instead, both transmitted and reflected contributions saturate near unity, despite the absence of lifted degeneracy.

\begin{figure*}
\includegraphics[width=\textwidth,height=0.8\textheight,keepaspectratio]{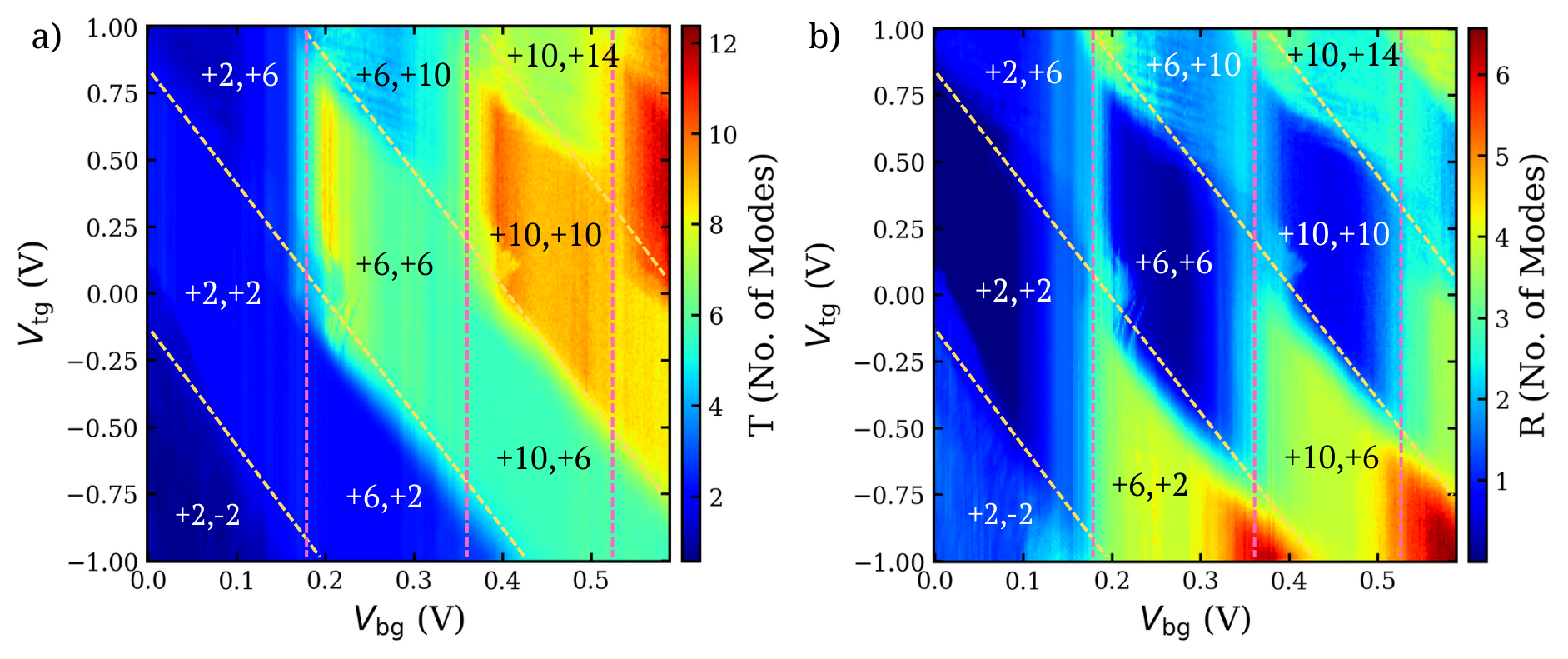}
\caption{\label{fig3}\textbf{The experimental results obtained from a surface plot of conductance:} a)transmission and b)reflection. Fanlines corresponding to the filling factors of the backgate and QPC are represented by pink and yellow dashed lines, respectively. The filling factors corresponding to the backgate and QPC region are marked. The data is obtained after applying Landauer-Buttiker Formalism over the QPC device to translate the raw voltage values into the transmitted and reflected modes.}
\end{figure*}

For positive top-gate voltages ($V_{\mathrm{QPC}}>0.1$~V), the local filling beneath the gate increases from $\nu_\mathrm{tg}=2$ toward $\nu_\mathrm{tg}=6$. Although additional localized states are expected to remain decoupled from the propagating edge channel, we observe a strong suppression of transmission, reducing the transmitted mode number to $N=1$. To elucidate this behavior, we perform two-gate spectroscopy to map the transmitted and reflected modes as functions of $\nu_\mathrm{bg}$ and $\nu_\mathrm{tg}$ (Figs.~\ref{fig3}a,b).

The resulting maps reveal a reduction in transmitted modes whenever $\nu_\mathrm{tg}\neq\nu_\mathrm{bg}$, most pronounced for the bipolar region . In the unipolar regime, for example, at $\nu_\mathrm{bg}=6$, tuning $\nu_\mathrm{tg}$ from 6 to 2 yields $N=2$ transmitted and $M-N=4$ reflected modes. In the bipolar regime, at $\nu_\mathrm{bg}=2$, sweeping $\nu_\mathrm{tg}$ from 2 to $-2$ increases the reflected contribution from $M-N=0$ to $\sim1.2$, while the transmitted mode is reduced to $N\sim0.8$. The appearance of an effectively single transmitted mode is unexpected in view of the degeneracy of quantum Hall edge states. We attribute this behavior, in both regimes, to enhanced mode mixing between propagating edge channels and localized states beneath the top gate.


\section{\label{sec:model}Microscopic Model and Numerical Methods}

We model the graphene QPC in a non-interacting single-particle tight-binding framework using the \textsc{kwant} package \cite{Groth2014}. For computational efficiency, we use the scalable model introduced in Ref.\cite{Ricthcer2015}. By computing the spatially varying potential profile generated by metallic gates above a two-dimensional electronic system, we mapped the top-gate and back-gate voltages onto the graphene sheet to produce the device's electrostatic potential landscape. Previous works mostly use a rectangular potential or a saddle point like potential profile to simulate QPC effects, but these are not the realistic effects which would be seen in experiments. This procedure accounts for the finite geometry of the gates and yields a realistic description of the confinement within the QPC region. Hence, the model captures electrostatics and lattice-scale physics simultaneously. 
The carrier density induced by the electrostatic gates is modeled within a parallel-plate capacitor approximation. Contributions from the backgate and top gate are given by
$
n_{\mathrm{bg}} = \varepsilon_0 \varepsilon_r V_{\mathrm{bg}}/e d_{\mathrm{bg}
},
$ and $
n_{\mathrm{tg}} = \varepsilon_0 \varepsilon_r V_{\mathrm{tg}}/e d_{\mathrm{tg}
} ,
$
where $\varepsilon_0$ is the vacuum permittivity, $\varepsilon_r$ is the relative dielectric constant of the gate dielectric, $e$ is the elementary charge, and $d_{\mathrm{bg}}$ ($d_{\mathrm{tg}}$) denotes the distance between the graphene sheet and the backgate (top gate). The applied gate voltages are denoted by $V_{\mathrm{bg}}$ and $V_{\mathrm{tg}}$, respectively.
The total local carrier density is taken as the sum of the two gate-induced contributions, $ n_{\mathrm{net}} = n_{\mathrm{bg}} + n_{\mathrm{tg}}$ .

In graphene, the carrier density is related to the Fermi energy through the linear Dirac dispersion. The local Fermi energy corresponding to a carrier density $n$ is given by $E_F = \mathrm{sgn}(n)\,\hbar v_F \sqrt{\pi |n|}$ ,
where $v_F$ is the Fermi velocity of graphene and $\mathrm{sgn}(n)$ accounts for electron ($n>0$) and hole ($n<0$) doping.

Using this relation, the spatially varying onsite potential entering the tight-binding Hamiltonian is obtained directly from the local carrier density profile $n(x,y)$ generated by the gate electrostatics,
\begin{equation}
V(x,y) =
-\,\mathrm{sgn}[n(x,y)]\,\hbar v_F \sqrt{\pi |n(x,y)|} .
\end{equation}
This mapping ensures that the onsite energy accurately reflects the local electrochemical potential landscape imposed by the gates, including both the magnitude and polarity of the charge carriers.

The resulting tight-binding Hamiltonian describing the graphene QPC is written as
\begin{equation}
H =
\sum_i V(x_i,y_i)\, c_i^\dagger c_i
- t \sum_{\langle i,j \rangle} c_i^\dagger c_j ,
\end{equation}
where $c_i^\dagger$ ($c_i$) creates (annihilates) an electron on lattice site $i$, $t$ is the nearest-neighbor hopping amplitude, and $\langle i,j \rangle$ denotes nearest-neighbor pairs. This Hamiltonian provides a lattice-resolved description of charge transport through the electrostatically defined QPC.

A perpendicular magnetic field is incorporated via the Peierls substitution using the Landau gauge, whereby the field enters as a phase factor in the hopping amplitudes, ensuring gauge-invariant coupling. All simulations are performed in parameter regimes where the magnetic length remains well resolved relative to the lattice discretization and system size.

In the presence of a perpendicular magnetic field, the graphene energy spectrum is quantized into Landau levels with energies $E_{N} = \mathrm{sgn}(N_L)\, v_F \sqrt{2 e \hbar B |N_L|}, N_L \in \mathbb{Z}$, reflecting the relativistic dispersion of graphene and defining the filling-factor structure relevant for quantum Hall transport.

To obtain the spectral properties of the system, we employ the Kernel Polynomial Method (KPM) to evaluate the spectral density of the isolated scattering region. The density of states is formally defined as
\begin{equation}
\rho(E) =
\mathrm{Tr}\!\left[ \delta(E - H) \right] ,
\end{equation}
where $H$ is the tight-binding Hamiltonian of the scattering region. Within KPM, the Dirac delta function is approximated by a finite expansion in Chebyshev polynomials.
\begin{equation}
\delta(E - H)
\approx
\sum_{m=0}^{M_n}
\mu_m\, g_m\, T_m(E) ,
\end{equation}
where $T_m(E)$ are Chebyshev polynomials of the first kind, $\mu_m$ are the corresponding expansion moments, $g_m$ are kernel coefficients, and $M_n$ the number of retained moments. Convergence was systematically verified with respect to both $M_n$ and magnetic field strength.

\begin{figure*}
\includegraphics[width=\textwidth,height=0.8\textheight,keepaspectratio]{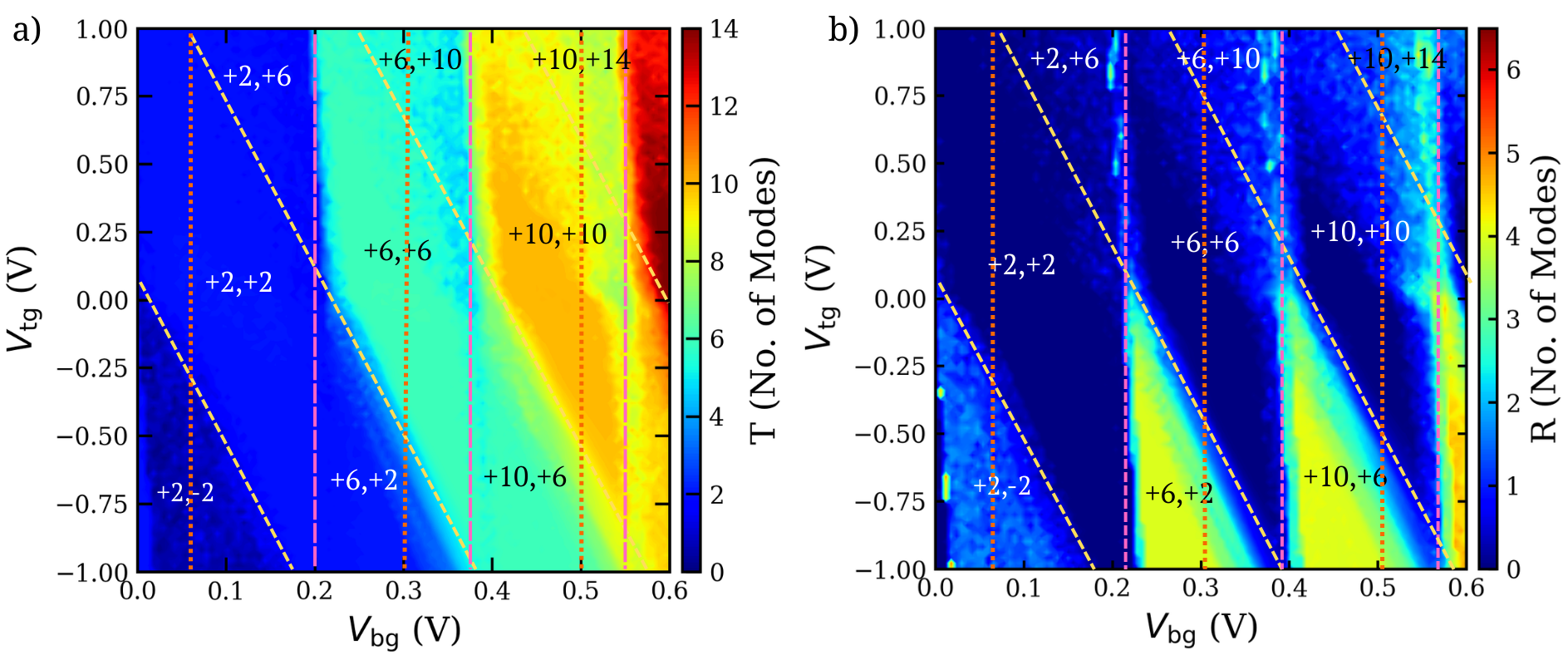}
\caption{\label{fig4} \textbf{KWANT tight-binding simulation:} The resulting a) Transmission and b) Reflection across the QPC using KWANT tight-binding simulation. The fanlines of banckgate and QPC are represented by pink and yellow dashed lines. The filling factor of the backgate and QPC are also shown at different regions. The orange dotted line showcases the linecuts taken for further analysis in Figure 5.}
\end{figure*}

The carrier density is obtained by integrating the density of states up to a given energy, $n(E) = \frac{2}{A} \int_{\infty}^{E}
\rho(E')\, dE' $, where the factor of $2$ accounts for spin degeneracy and $A$ is the area of the scattering region. Inverting this relation establishes a direct correspondence between the local carrier density and the onsite potential imposed by the electrostatic gates, ensuring consistency between the gate voltages and the resulting filling factors.

Random onsite potential fluctuations of tunable strength are added to the electrostatic landscape to model disorder phenomenologically. This disorder captures the effects of charge puddles, substrate inhomogeneity, and fabrication-induced potential fluctuations present in experimental devices, enabling realistic simulations of transport through the QPC. Disorder is implemented using the uniform disorder model in \textsc{kwant}, and all results are averaged over multiple disorder realizations. The resulting Hamiltonian is used to compute transmission probabilities and two-terminal conductances within the Landauer–Büttiker formalism, as well as the number of transmitted and reflected modes, spatial current distributions, and shot noise.

\section{\label{sec:res}RESULTS}

\subsection{\label{sec:edge}Edge-Channel Transmission Through the QPC}

We evaluate the two-terminal transmission and reflection using the tight-binding model with geometry similar to that of experimental device (see Methods). The simulations reproduce the quantized quantum Hall plateaus and the characteristic slopes associated with the back-gate and QPC filling factors, as shown in Fig.~\ref{fig4}, enabling direct comparison with experiment and providing microscopic insight into edge-channel scattering at the constriction.

Figure~\ref{fig3} presents experimentally extracted maps of the transmitted mode number $N$ [panel~(a)] and the reflected mode number $M - N$ [panel~(b)] as functions of back-gate voltage $V_\mathrm{bg}$ and top-gate voltage $V_\mathrm{tg}$ at a fixed magnetic field $\textrm{B} = 10$ T. The color scale indicates the number of quantum Hall edge channels transmitted through or reflected by the QPC. Well-defined plateaus of nearly constant transmission and reflection appear, separated by sharp transitions. The dashed diagonal lines trace contours of constant filling factor in the bulk and in the QPC region (the confined area defined by the back and top gates). Their slopes reflect the relative capacitive coupling of the two gates to the graphene sheet. 

Naively, quantum transport in a Hall bar with a QPC would be expected to depend primarily on the filling factor within the confined QPC region. However, the transport behavior is governed more generally by the interplay between the filling factors $\nu_\mathrm{bg}$, $\nu_\mathrm{tg}$, and $\nu_\mathrm{QPC}$, as well as their respective polarities. Here, $\nu_\mathrm{bg}$, $\nu_\mathrm{tg}$, and $\nu_\mathrm{QPC}$ denote the filling factors in the bulk, beneath the top gate, and within the constriction, respectively.

At fixed back-gate voltage (dotted orange line in Fig.~\ref{fig4}) $V_\mathrm{bg}$, sweeping the top-gate voltage $V_\mathrm{tg}$ gives rise to three distinct regimes:
\begin{enumerate}
\renewcommand{\labelenumi}{(\roman{enumi})}
\item $|\nu_\mathrm{tg}| \leq |\nu_\mathrm{QPC}| \leq |\nu_\mathrm{bg}|$,
\item $|\nu_\mathrm{tg}| \approx |\nu_\mathrm{QPC}| \approx |\nu_\mathrm{bg}|$,
\item $|\nu_\mathrm{tg}| \geq |\nu_\mathrm{QPC}| \geq |\nu_\mathrm{bg}|$.
\end{enumerate}

Each regime can further be classified according to the polarity of the top- and back-gate filling factors: a unipolar regime ($\nu_\mathrm{tg}$ and $\nu_\mathrm{bg}$ have the same sign) and a bipolar regime ($\nu_\mathrm{tg}$ and $\nu_\mathrm{bg}$ have opposite signs). The filling factor in the QPC is determined by the competition between the electrostatic potentials induced by the top and back gates.

In the two-gate spectroscopy shown in Figure~\ref{fig4}, we first examine the diamond-shaped regions where $|\nu_\mathrm{bg}| = |\nu_\mathrm{QPC}| = |\nu_\mathrm{tg}|$ (case (ii)). In this regime, the electrostatic potential is approximately uniform across the device, allowing the quantum Hall edge channels to propagate adiabatically through the QPC without significant potential mismatch.

The specific trajectories of the incoming edge modes depend on the polarity of the filling factor beneath the top gate, $\nu_\mathrm{tg}$. In the unipolar regime, where $\nu_\mathrm{tg}$ has the same sign as the bulk filling factor, the edge states pass beneath the top gate and are fully transmitted through the constriction. In contrast, in the bipolar regime, where $\nu_\mathrm{tg}$ has opposite polarity, a $p$–$n$ interface forms within the QPC. The edge modes are then redirected through the confined region of the constriction, where they interact with counter-propagating states, leading to finite scattering.

This picture is consistent with the near-zero reflection observed in Figs.~\ref{fig3} and \ref{fig4}(b), corresponding to the dark-blue diamond-shaped regions. In the bipolar regime, propagation of higher Landau levels into the top-gated region is strongly suppressed, and transport across the QPC becomes confined to the narrow constriction.

We now turn to the regime where $\nu_\mathrm{bg}$ is fixed while $\nu_\mathrm{QPC}$ is reduced (case (i): $\nu_\mathrm{tg} \leq \nu_\mathrm{QPC} \leq \nu_\mathrm{bg}$). In this situation, the number of edge channels injected from the bulk exceeds the number of modes supported by the constriction. The QPC therefore acts as a mode filter: only $\nu_\mathrm{QPC}$ channels are transmitted, while the excess $\nu_\mathrm{bg}-\nu_\mathrm{QPC}$ modes are reflected. This selective transmission produces quantized plateaus in the transmitted mode number $N$, accompanied by a complementary increase in the reflected modes $M-N$.

In the unipolar regime, increasing the filling factor beneath the top gate introduces additional edge channels within the constriction. The transmitted mode can then hybridize with localized states under the top gate, leading to mode mixing. As $V_\mathrm{tg}$ is increased further (case (iii)), this mixing becomes more pronounced, and the transmission progressively deviates from the ideal quantum Hall plateau. The quantized plateaus consequently weaken, giving way to partial transmission and reflection.

A qualitatively different behavior emerges in the bipolar regime. When the filling factor beneath the top gate has opposite polarity to that of the bulk, a $p$–$n$ junction forms inside the constriction. For the zeroth Landau level, the edge states follow snake-like trajectories along this interface. Transport across the QPC is then governed by resonant tunneling via localized states beneath the top gate, resulting in a strongly oscillatory transmission that can approach unity. This distinct response further supports the interpretation based on edge-mode trajectories for different Landau-level fillings in the conductance map, as discussed below.

\subsection{\label{sec:mixing}Microscopic Origin of Edge-Channel Mixing}

\begin{figure*}
\includegraphics[width=\textwidth,height=0.8\textheight,keepaspectratio]{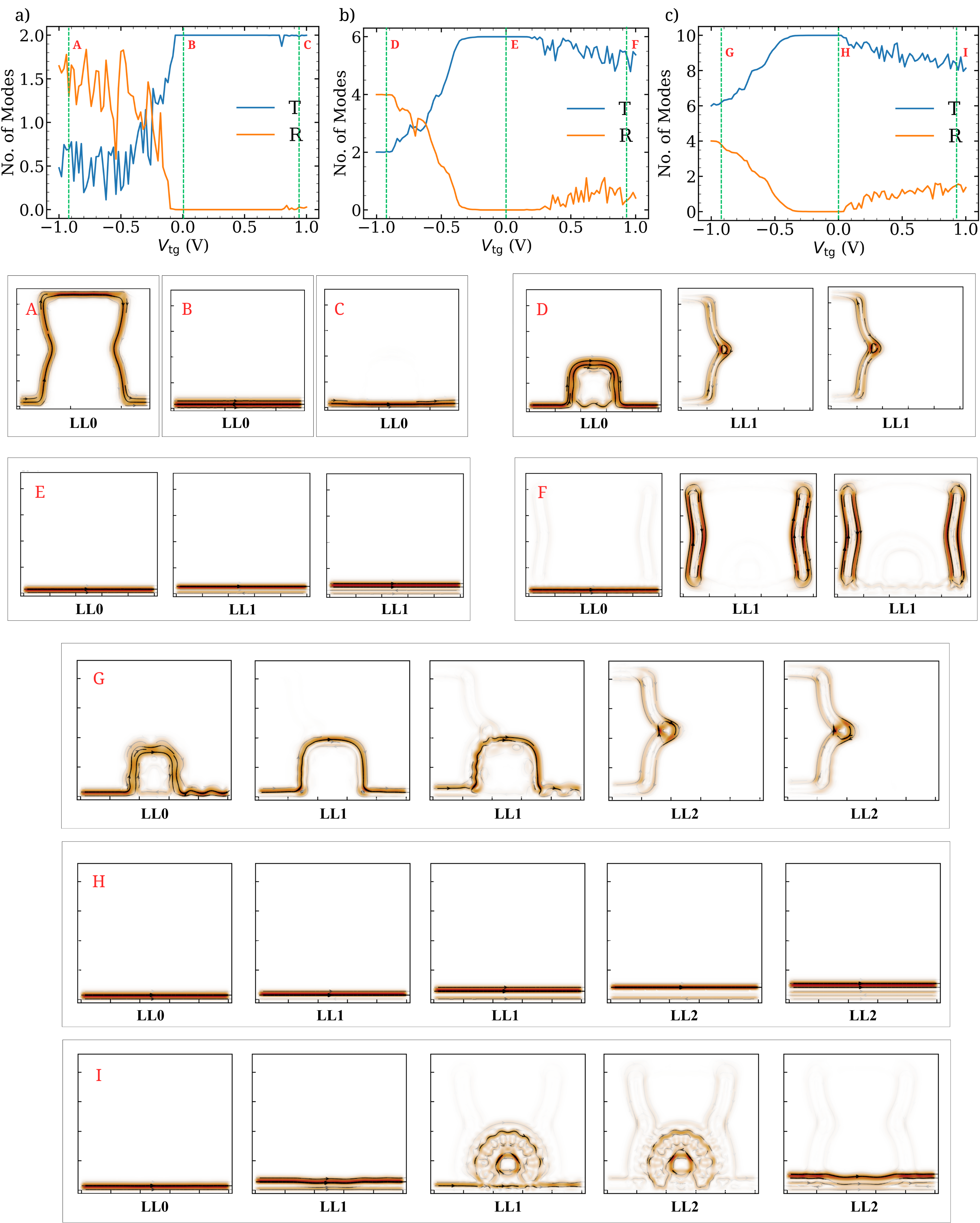}

\caption{\label{fig5}\textbf{The simulated mode trajectories obtained from different regimes:} Line cuts of the transmission (solid blue) and reflection (dashed orange) as a function of top gate voltage for three representative backgate voltages: (a) $V_\mathrm{bg}=0.05~\mathrm{V}$ ($\nu_\mathrm{bg}=2$), (b) $V_\mathrm{bg}=0.3~\mathrm{V}$ ($\nu_\mathrm{bg}=6$), and (c) $V_\mathrm{bg}=0.5~\mathrm{V}$ ($\nu_\mathrm{bg}=10$). Vertical dashed lines mark selected gate voltages at which real-space current density distributions are evaluated. Panels A--I show the corresponding mode-resolved current density maps for the indicated gate voltages. Panel A illustrates the propagation of the zeroth Landau level $N_L= 0$ in the bipolar regime. Panel B corresponds to the case $\nu_\mathrm{bg} = \nu_\mathrm{QPC} = \nu_\mathrm{tg}$, where the electrostatic potential is approximately uniform across the constriction, and the zeroth Landau level is transmitted adiabatically under the top gate without reflection. Panel C shows the continued adiabatic transmission of edge modes. Panels D--F correspond to the constant backgate filling-factor regime at $\nu_\mathrm{bg} = 6$, where additional incoming edge channels are injected toward the constriction defined by the top gate potentials. In this regime, the QPC acts as a mode filter: only a fixed number of modes are transmitted through the constriction, while the excess modes are reflected, as evidenced by the spatial separation of transmitted and reflected trajectories. Panel I illustrate the unipolar high-density regime under the top gate. Here, multiple edge channels coexistunder the topgate and within the constriction. In this case, the potential landscape becomes increasingly non-adiabatic, leading to enhanced mode mixing and partial transmission of individual channels, consistent with the degradation of conductance plateaus observed in the line cuts. The detailed classification of the different modes of transmission is presented in the table \ref{tab1}.}
\end{figure*}

To pinpoint where mode mixing occurs in the graphene QPC, we analyze the spatial current density for selected points along the line cuts in Fig.\ref{fig5}(a-c), covering the configurations labeled A-I and summarized in Table \ref{tab1}. These snapshots show how the relative position and overlap of edge trajectories evolve with $\nu_\mathrm{bg}, \nu_\mathrm{tg}$, and $\nu_\mathrm{QPC}$ and thereby distinguish regimes of mode filtering, adiabatic transport, and multi-mode mixing.

In the bipolar configuration at the lowest Landau level (region A), only the zeroth-Landau-level edge channel maintains a continuous path through the constriction. The current density reveals a single narrow interface trajectory that tunnels between opposite edges along the $p$-$n$ boundary, without visible branching into simultaneously transmitted and reflected components. Despite the nontrivial conductance features, this pattern corresponds to effectively single-mode transport with minimal inter-channel mixing in the constriction itself.

In the regime of $|\nu_\mathrm{bg}| \ge |\nu_\mathrm{tg}, \nu_\mathrm{QPC}|$ (regions D and G), several edge channels impinge on the QPC, but only a subset can be guided under the top gate. The current-density maps show that some modes follow smooth trajectories beneath the split gates, while the remaining channels are cleanly diverted around the gated region and reflected along the opposite edge. Transmitted and reflected paths are spatially well separated, so each incoming mode behaves almost dichotomously, either nearly fully transmitted or nearly fully reflected, which is characteristic of mode filtering rather than mixing. However, it is possible to partially reflect the incoming modes; see Supplementary Fig.~7. Interestingly, this partial reflection mimics the signature of the quantized plateaus at the lifted degenerate values in the transmitted N and reflected modes M-N, which may be related to disorder or device geometry. This is warranted for further investigation.
Furthermore, mode mixing becomes pronounced when the top-gate filling exceeds the bulk filling, $|\nu_\mathrm{tg}, \nu_\mathrm{QPC}| > |\nu_\mathrm{bg}|$, corresponding to regions C, F, and I in Fig.~\ref{fig5}. In this regime, additional compressible and localized states develop beneath the split top gates and within the QPC, fundamentally modifying the character of edge transport.

In region C, the onset of this reconstruction is already visible: the edge trajectories begin to deviate from the clean spatial separation characteristic of mode filtering, and the inner channels approach localized states under the gate. The effect becomes much stronger in regions F and I. Here, multiple Landau-level edge channels are forced to co-propagate through the same narrow constriction, leading to strongly distorted and intertwined current paths.

While the outermost channel, particularly the zeroth Landau level, continues to follow a relatively smooth trajectory beneath the top gate, the inner channels repeatedly intersect localized states and scatter off the opposite device edge. The current-density maps clearly show branching of individual trajectories into transmitted and reflected components, evidencing partial back-scattering at the 
microscopic level.

This coexistence of extended propagating modes and gate-induced localized states is the key mechanism that transforms the QPC from a nearly ideal mode filter (bipolar regions: D and G) into a multi-mode scatterer (unipolar regions: F and I). Instead of being either fully transmitted or fully reflected, individual incoming channels are partitioned, acquiring transmission probabilities that take intermediate values between 0 and 1. This microscopic partitioning directly explains the degradation of conductance plateaus and the emergence of enhanced shot noise, discussed in the following section.

The matched-filling configurations $|\nu_\mathrm{bg}| = |\nu_\mathrm{tg}| = |\nu_\mathrm{QPC}|$ (regions B, E, and H) serve as a reference where mixing is minimal. Here, both lowest and higher Landau-level edge channels follow nearly parallel, non-crossing trajectories beneath the top gates, with no pronounced approach to localized states or to the opposite edge. The absence of branching in these maps indicates that scattering between modes in the constriction is weak and that the QPC operates close to the adiabatic limit.

In summary, mode mixing in this graphene QPC is confined to gate configurations where several edge channels are forced to co-propagate through a non-adiabatic, electrostatically reconstructed region containing localized states under the top gates. In these regions, the microscopic partitioning of individual modes into transmitted and reflected branches sets the stage for enhanced partition noise and elevated Fano factors, which we analyze quantitatively in the following section.
\begin{table}
  \centering
  \setlength{\tabcolsep}{6pt} 
  \resizebox{0.4\textwidth}{!}{%
    \begin{tabular}{|c|c|c|}
      \hline
      Mode & Adiabatic & Mode\\
      Filtering & Transmission & Mixing \\
      $\nu_\mathrm{bg}\geq \nu_\mathrm{tg}, \nu_\mathrm{QPC}$ & $\nu_\mathrm{bg} = \nu_\mathrm{tg} = \nu_\mathrm{QPC}$ & $\nu_\mathrm{bg}\leq \nu_\mathrm{tg}, \nu_\mathrm{QPC}$\\
      \hline\hline
      \textbf{A} & \textbf{B} & \textbf{C} \\
      (2,~-2) & (2,~2) & (2,~6)\\
      \hline
      \textbf{D} & \textbf{E} & \textbf{F} \\
      (6,~2) & (6,~6) & (6,~10) \\
      \hline
      \textbf{G} & \textbf{H} & \textbf{I}\\
      (10,~6) & (10,~10) & (10,~14) \\
      \hline
    \end{tabular}%
  }
  \caption{Table 1: Depicting three main regions based on mode trajectories across the QPC. Here regions marked as \textbf{A, B....I} in the Fig.\ref{fig5}a, b, and c. Each cell of the table carries the filling due to the back gate and in the QPC region ($\nu_\mathrm{bg}$, $\nu_\mathrm{QPC}$).}
  \label{tab1}
\end{table}

\subsection{\label{sec:shotnoise}Shot Noise and Fano Factor}

Having established that the model accurately reproduces the experimentally observed conductance maps, we now exploit this validated framework to compute the two-terminal shot noise associated with transport through the QPC. Shot noise, arising from the discrete nature of charge carriers, offers a complementary observable that is sensitive to mode partitioning and quantum fluctuations at the constriction.

In the quantum Hall regime, shot noise vanishes when edge channels are either fully transmitted ($T = 1$) or fully reflected ($T = 0$), showing the absence of partitioning. Finite shot noise, therefore, directly signals partial transmission and mode mixing at the constriction. The Landauer formula gives the shot noise power at zero temperature \cite{Blanter2000}:
\begin{equation}
S = \frac{2e^2}{h}V \sum_n T_n(1-T_n)
\label{Fano}
\end{equation}
where $V$ is the applied bias voltage and $T_n$ are the transmission eigenvalues of the individual channels. The Fano factor $F = S/2eIG$, where $G$ is the conductance, provides a dimensionless measure of shot noise that characterizes the statistics of transmission.

\begin{figure*}
\includegraphics[width=\textwidth,height=0.8\textheight,keepaspectratio]{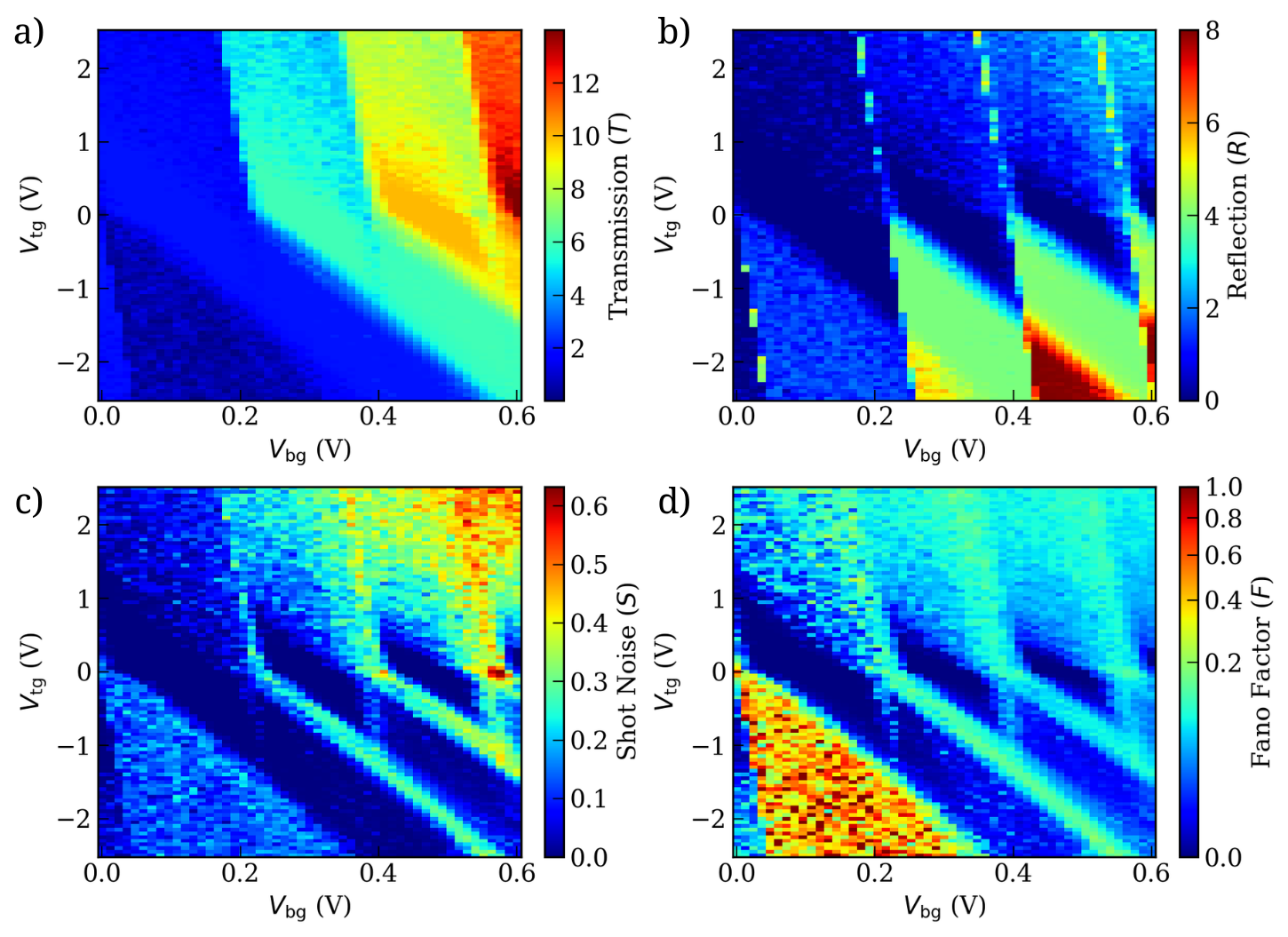}
\caption{\label{fig6}\textbf{The simulated Conductance, Shot noise and Fano factor map:} Comparison of conductance, shot noise, and Fano factor across gate-defined transport regimes of the graphene quantum point contact. (a) Two-terminal conductance map $G$ as a function of backgate voltage ($V_{\mathrm{bg}}$) and top gate voltage ($V_{\mathrm{tg}}$), showing quantized plateaus and transition regions corresponding to different numbers of transmitted quantum Hall edge modes. (b) Corresponding shot noise map evaluated within the Landauer--Büttiker formalism, where finite noise arises from the partitioning of edge channels at the constriction. (c) Fano factor $F = S/G$, characterizing the degree of mode partitioning and quantum fluctuations relative to Poissonian transport. In regions of quantized conductance plateaus, where edge channels are either fully transmitted or fully reflected, the shot noise is strongly suppressed and the Fano factor approaches zero, reflecting the absence of partitioning. In contrast, enhanced shot noise and elevated Fano factor are observed along the transition lines between plateaus, where modes are partially transmitted through the quantum point contact. Notably, the bipolar regime exhibits pronounced shot noise and large Fano factor values despite reduced conductance, indicating strong partitioning associated with tunneling of the zeroth Landau level across the $p$--$n$ interface. These results demonstrate that shot noise and the Fano factor provide complementary information to conductance measurements, revealing microscopic mode partitioning and quantum fluctuations that are not directly accessible through conductance alone.}
\end{figure*}

Figure~\ref{fig6} displays the calculated conductance, shot noise, and Fano factor as functions of back-gate and top-gate voltages, highlighting how mode partitioning evolves across different transport regimes.

In the unipolar regime at fixed $\nu_{\mathrm{bg}}$, increasing the top-gate voltage progressively enhances mode mixing within the constriction. This is directly reflected in the increase of shot noise, as individual edge channels become partially transmitted. By contrast, within the conductance plateaus, where each mode is either fully transmitted or fully reflected, the shot noise vanishes, consistent with transmission eigenvalues $T_n \approx 0$ or $1$ and the absence of partitioning.

In the bipolar regime, where the QPC acts as a mode filter across a gate-defined $p$–$n$ interface, the shot noise exhibits a pronounced modulation between nearly zero and finite values. These oscillations correlate closely with the reflected current (see Supplemental Figs.~S2 and S3), indicating alternating regimes of weak and strong partitioning. The rapid enhancement of shot noise and consequently of the Fano factor $F$ originates from partial reflection of edge channels as the transmission deviates from unity. Since $F \propto \sum_n T_n(1-T_n)/\sum_n T_n$, decreasing total transmission amplifies the relative contribution of partition fluctuations, leading to large $F$ in the limit $T \to 0$.

\begin{figure*}
\includegraphics[width=\textwidth,height=0.8\textheight,keepaspectratio]{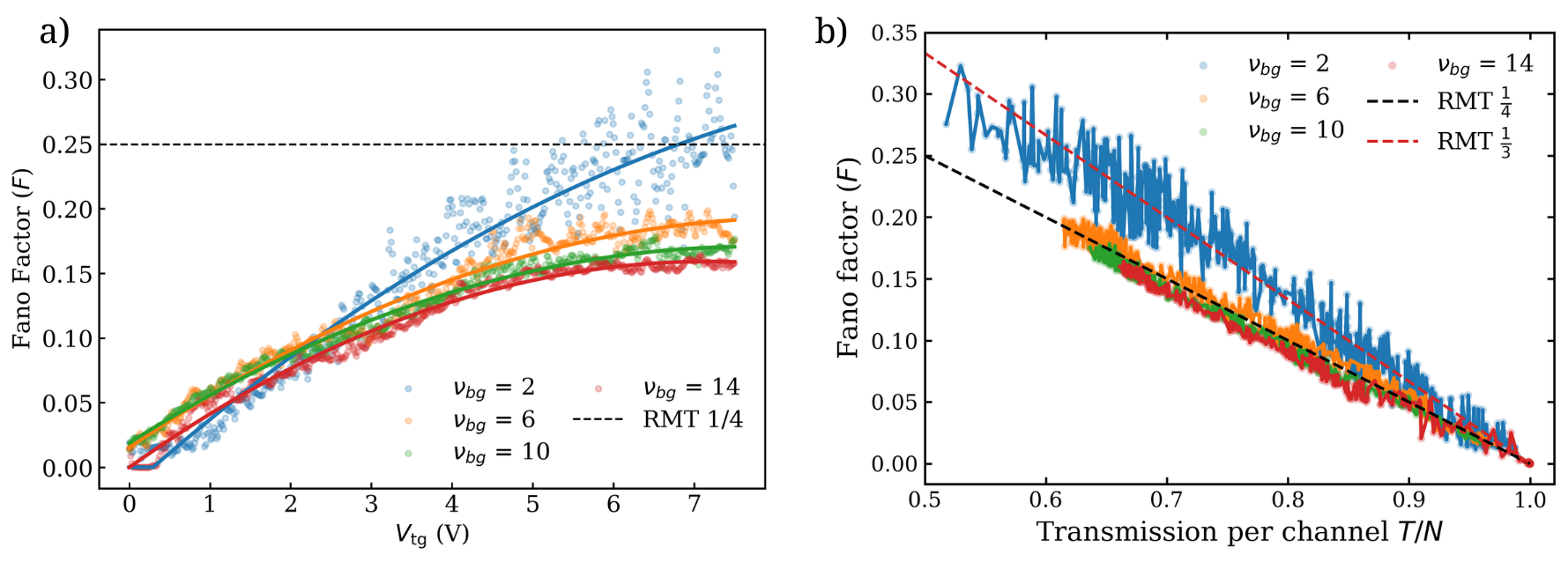}
\caption{\label{fig7} \textbf{Fano factor evolution and convergence to universal random matrix theory values:} (a)~Fano factor $F$ as a function of top-gate voltage $V_{\mathrm{tg}}$ for the zeroth Landau level ($N_L=0$, $\nu_{\mathrm{bg}}=2$, blue) and higher Landau levels ($N_L>0$, $\nu_{\mathrm{bg}}$=6 (orange), 10 (green), 14 (red)). Solid lines show numerical fitting, yielding asymptotic values $F_\infty^{(N_L=0)} \simeq 0.33$ and $F_\infty^{(N_L=1)} \simeq 0.254$.  The theoretical RMT predictions for a chaotic cavity ($F = 1/4 = 0.250$, dash-dotted), derived in Appendix A, are shown for reference. The fitted values approach, but do not exactly reach the RMT limits due to finite QPC length and incomplete mode mixing. Increasing disorder potential will result in quicker saturation of the Fano factor to the chaotic limit. 
(b)~Fano factor $F$ versus transmission per mode $T/N$ for both Landau level families. As transmission increases $T\to1$, $F$ saturates toward the ballistic limit $F\to0$, consistent with the Landauer formula [Eq.~\ref{Fano}]. The $N_L = 0$ and $N_L > 0$ data separate at intermediate transmission, reflecting the different transmission eigenvalue distributions 
(peaked near $T = 1$ for $N_L = 0$ vs.\ bimodal for $N_L > 0$) that give rise to the distinct Fano factor values in panel~(a).
        }
\end{figure*}

We now focus on the evolution of the Fano factor in the unipolar region with an increase in the number of modes under the top gate.
In Fig.~\ref{fig7}(a), the Fano factor $F$ converges to a distinct asymptotic limit as the top-gate voltage increases. In particular, the saturation limit of the zeroth Landau level differs significantly from that of the higher Landau levels.
We observe a linear dependence between conductance $G$,  Fano factor $F$, and average transmission per channel $\langle T \rangle/N$. We observe that the fitted line for higher Landau levels falls at a Fano factor of $F = 1/4$ when the transmission per channel $\langle T \rangle/N = 1/2$. While a Fano factor of 1/4 is also characteristic of asymmetric transport through a small number of quasi-ballistic modes, for example, for transmissions $T_1 \sim 0.15$ and $T_2 \sim 0.85$ of two degenerate channels of a Landau level will give rise to $F \simeq 0.255$ \cite{Kumar2012}. However, it is non-universal and strongly gate dependent. 
In contrast to this, our calculation converges to the universal value of $F = 1/4$ characteristic of chaotic cavities \cite{Oberholzer2001, Savin2006}. As detailed in Appendix A, this universality arises from complete mode mixing and standard random-matrix statistics involving many channels under the top gate.
For the bimodal distribution of transmission eigenvalues characteristic of standard chaotic mode mixing ($N_L > 0$), the statistical averages result in \cite{Martin2003}:
\begin{equation*}
\langle T \rangle = \frac{1}{2}, \qquad
\langle T(1 - T) \rangle = \frac{1}{8},
\end{equation*}
and hence
\begin{equation*}
F = \frac{\langle T(1 - T) \rangle}{\langle T \rangle}
= \frac{1/8}{1/2}
= \frac{1}{4}.
\end{equation*}
This result is entirely independent of the microscopic details or the specific shape of the scattering potential. Strikingly, for the zeroth Landau level ($N_L = 0$), the Fano factor saturates to $F \simeq 1/3$. While this value has previously appeared in analytical works on zero-field graphene and is conventionally attributed to Dorokhov-like transmission-eigenvalue distributions arising from diffusive bulk transport~\cite{Tworzydlo2006, Danneau2008, Schuessler2010}, our tight-binding approach makes no such assumption. The model reproduces the experimentally observed conductance through purely elastic scattering without invoking diffusive physics, and our computational results for the noise therefore serve as a firm numerical benchmark independent of existing phenomenological labels.

\par The RMT framework developed in Appendix~A provides the 
correct microscopic interpretation and a unifying explanation for both noise limits. Transport through the $N_L = 0$ channel is governed by single-channel ($N = 1$) chaotic mixing, for which the transmission eigenvalue distribution is exactly flat, $\rho(T) = 1$, yielding $F = 1/3$ exactly. This mechanism is physically distinct from the Dorokhov scenario: the same numerical value arises from a fundamentally different eigenvalue 
distribution, not from disorder-driven diffusion but from the sublattice polarization of the $N_L = 0$ state, which suppresses coupling to the mixed-sublattice localized states beneath the gate and confines transport to an effective single channel. For higher Landau levels, complete multi-channel mixing ($N \gg 1$) yields the bimodal eigenvalue distribution and $F = 1/4$. The power of this framework lies in its independence from microscopic scattering parameters: once the channel number and mixing probability are identified, the noise statistics follow universally from the eigenvalue distribution alone~\cite{Beenakker1997, BeenakkerButtiker1992}.

\section{\label{sec:conc}CONCLUSION}
We have developed a hybrid theoretical framework that unifies device-realistic tight-binding simulations of gate-defined graphene quantum point contacts with random-matrix theory to predict and interpret shot-noise signatures across distinct quantum Hall transport regimes. 
The transmission eigenvalue spectra, validated against mode-resolved conductance maps in hBN-encapsulated graphene Hall bars reveal three microscopically distinct regimes: adiabatic propagation at matched filling, sharp mode filtering when the QPC acts as a selective barrier, and multi-mode mixing driven by localized states beneath the split gate when the local filling factor exceeds the bulk.

Our central result is a Landau-level-resolved crossover in the Fano factor. For higher Landau levels ($N_L > 0$), complete mode mixing within an electrostatically reconstructed cavity involving many channels produces a universal value $F \simeq 1/4$, consistent with the random-matrix prediction for a chaotic scatterer in the macroscopic limit $N \gg 1$, where the transmission eigenvalue distribution is bimodal with peaks at $T \to 0$ and $T \to 1$. In contrast, the zeroth Landau level ($N_L = 0$) yields $F \simeq 1/3$. This value has a precise RMT origin: transport through the $N_L = 0$ channel is governed by single-channel ($N = 1$) chaotic mixing, for which the transmission eigenvalue distribution is exactly flat, yielding $F = 1/3$ exactly. This is mechanistically distinct from, though numerically coincident with, the pseudo-diffusive $F = 1/3$ known from zero-field graphene near the Dirac point, where the same value arises from a Dorokhov-like distribution generated by evanescent bulk modes \cite{Tworzydlo2006, Danneau2008}. The distinction reflects the unique character of the $N_L = 0$ quantum Hall state: its particle--hole-symmetric, sublattice-delocalized structure confines transport to effectively a single mixed channel even in the quantum Hall regime, a property absent in all higher Landau levels.

The resulting $F = 1/3$ versus $F = 1/4$ crossover provides a universal discriminator between chaotic and pseudo-diffusive edge transport in graphene QPCs. Beyond a mere quantitative change in noise level, it represents a Landau-level-resolved transition between two distinct transport universality classes within a single mesoscopic device. This behavior differs fundamentally from that of conventional semiconductor heterostructure QPCs, where only the chaotic-cavity regime is accessible; graphene QPCs in the quantum Hall regime uniquely host both universality classes simultaneously, with the Landau-level index acting as a discrete, gate-tunable selector. This is a direct consequence of the relativistic band structure of graphene and, to our knowledge, has no analogue in non-Dirac two-dimensional systems.

These results have direct implications for transport experiments in graphene Hall-bar devices. It provide concrete experimental benchmarks for noise spectroscopy in graphene Hall-bar devices. The $F \simeq 1/4$ limit is universal: once several Landau levels are forced through an electrostatically reconstructed constriction, gate geometry, disorder configuration, and precise filling factors all become irrelevant--the noise is governed entirely by the chaotic-cavity ensemble. The $F \simeq 1/3$ limit at $N_L = 0$ is equally robust, reflecting the single-channel character imposed by sublattice polarization rather than bulk diffusion. Deviations from either benchmark would constitute direct experimental signatures of incomplete mixing, residual coherence, or inelastic relaxation.

Our work complements recent experiments on enhanced shot noise in graphene QPCs with electrostatic reconstruction~\cite{Garg2025}, which focuses on correlated tunneling, by providing universal noise benchmarks $F \simeq 1/3$ (single-channel, $N_L = 0$) versus $F \simeq 1/4$ (chaotic, $N_L > 0$) that are robust against microscopic details. The experimental observation of this crossover would constitute a direct test of the distinct transport universality classes accessible within a single device via Landau-level tuning and noise spectroscopy at biases well above the thermal floor, $eV \gg k_{\mathrm{B}}T$, in the same hBN-encapsulated Hall-bar geometry.

Beyond graphene QPCs, our hybrid framework---combining device-level electrostatics with the statistical universality of random-matrix theory--- offers a general strategy for predicting noise signatures in gate-defined two-dimensional conductors where microscopic mode structure competes with universal scattering statistics. Natural extensions include bilayer graphene, where the zeroth Landau level has different sublattice character and the pseudo-diffusive regime is modified, moir\'{e} systems with flat-band Landau levels, and topological edge channels in quantum spin Hall devices. In this broader context, shot noise, long established as a probe of fractional charge in the fractional quantum Hall regime, emerges as an equally powerful tool for resolving the microscopic transport character of engineered integer quantum Hall constrictions.

\section{\label{sec:methods}Methods}

\subsection{Sample Fabrication}

h-BN-encapsulated graphene devices with graphite back gates were fabricated using a standard dry transfer technique. Graphene, h-BN, and graphite flakes were exfoliated onto 280 nm $\text{SiO}_2$/Si (p++) substrates. Suitable flakes were identified by optical microscopy, and monolayer graphene was confirmed by Raman spectroscopy. The selected crystals were sequentially picked using a PC/PDMS stamp to assemble an h-BN/graphene/h-BN/graphite heterostructure, which was then transferred to a 280 nm $\text{SiO}_2$/Si substrate (p++). The thickness of the top h-BN and bottom h-BN is approximately $25$ nm. Clean, bubble-free regions were identified by atomic force microscopy.

One-dimensional edge contacts were manufactured using a self-aligned process with PMMA as the e-beam resist, followed by reactive etching with $\text{CHF}_3/\text{O}_2$ to expose graphene edges and Ti/Au (5 nm/60 nm) metallization. The top split gates that define the QPC were patterned in a second lithography step using 20 keV e-beam lithography, followed by Ti/Au deposition, with a gate spacing of 100 nm. Contacts with the bottom graphite gate were defined simultaneously. Finally, Hall bar geometries were patterned and etched using $\text{CHF}_3/\text{O}_2$. The completed device had a typical size of ~$10$ $\mu$m $\times 5~ \mu$m .

\subsection{Simulation Parameters} 
\label{app:sim}

The tight-binding simulations were performed on a rectangular graphene sample of length $500$ nm and width $450$ nm. Two semi-infinite leads were attached to the transverse edges of the sample to model source and drain contacts. The electrostatic confinement defining the quantum point contact (QPC) was implemented using two top gates of width $100$ nm separated by a gap of $120$ nm. The top gates were positioned $25$ nm above the graphene layer, and the electrostatic potential was calculated by including a dielectric medium with relative permittivity $\varepsilon_r = 4$, corresponding to h-BN. A global back gate was placed $30$ nm below the graphene sheet, separated by the same dielectric constant, and was included in the potential profile.

To reduce computational cost while preserving the low-energy physics near the Dirac point, a scaling factor of $10$ was applied to the lattice, giving us an effective lattice constant of $a = 2.46 \text{nm}$. This procedure increases the effective lattice spacing while renormalizing the nearest-neighbor hopping parameter from its pristine value of $t = 2.7$ eV accordingly. For magnetic field simulations, a perpendicular field of $B = 3$ T was introduced via Peierls substitution. This corresponds to a magnetic length of $l_B \approx 14.8$ nm, which is larger than the effective lattice constant, and this imposes an upper energy constraint of $E_{\mathrm{max}} \ll 2.54$ eV for the validity of the scaling approach. Since the simulations probe energies very close to the Dirac point ($E = 10^{-6}$~eV), the Landau level physics in the relevant filling factor range is equivalent under the field rescaling, and the scaling does not affect the physical conclusions.

Disorder was incorporated into the KPM calculations through an onsite potential with a strength of 0.03 eV, representing static lattice disorder. In addition, random onsite disorder configurations were introduced to emulate realistic sample inhomogeneities. All reported results are averaged over five independent disorder realizations. This was done to match the experimental conductance map. Electronic transport properties were evaluated using numerical transmission and two-terminal shot noise calculations. From these quantities, the conductance, reflection probability, and shot noise were extracted and subsequently analyzed.

\section{Appendix}
\subsection{Random Matrix Theory for the shot noise}

\paragraph{Model.} 
We consider a system of $N$ spinless edge modes. The modes are transmitted under the top gate with probability $\Gamma$ without undergoing mode mixing. With probability $1-\Gamma$, they mix with the modes beneath the gate and are subsequently either transmitted to the opposite side of the sample or reflected. This mode mixing is modeled by a $2N \times 2N$ random unitary matrix drawn from the circular ensemble. Mathematically, this scattering problem is equivalent to two-terminal conduction through a chaotic quantum dot coupled to non-ideal leads with a tunnel probability of $1-\Gamma$, a system that has been analyzed extensively in previous works \cite{Brouwer1994, Brouwer1996}.

\begin{figure}[ht]
    \centering
    \includegraphics[width=0.9\columnwidth]{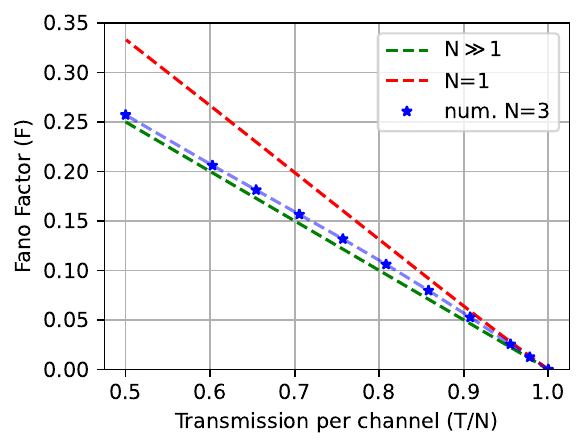}
    \caption{Fano factor $F$ as a function of the transmission per channel $\langle T \rangle/N$. The dashed curves indicate the analytical random-matrix theory predictions for a single channel ($N=1$) and the macroscopic limit ($N \gg 1$). Blue markers denote numerical simulations for $N=3$ modes. While the exact $N=1$ solution is markedly different, the numerical result for $N=3$ already converges to within $\sim 3\%$ of the $N \gg 1$ analytical limit.}
    \label{fig:fano_vs_transmission}
\end{figure}

\paragraph{Results: Average Transmission and Fano Factor.} 
In the macroscopic limit $N \gg 1$, the ensemble-averaged transmission $\langle T \rangle$ and the Fano factor $F$ can be evaluated in a closed analytical form by means of diagrammatic integration over the unitary group \cite{Brouwer1996}. Parameterizing in terms of the transmission probability $\Gamma$, we obtain $\langle T \rangle = N(1+\Gamma)/2$ and $F = (1-\Gamma)/4$. In Fig.\ \ref{fig:fano_vs_transmission}, we plot this analytical result by showing the dependence of $F$ on the average transmission per channel, $\langle T \rangle / N$. In the limit of complete mixing ($\Gamma \to 0$), we recover the universal random-matrix theory limit of $F=1/4$. For the single-channel case ($N=1$), an exact analytical solution can be derived from the algebraic formulas governing nonideal leads \cite{Brouwer1994}. As illustrated in Fig.\ \ref{fig:fano_vs_transmission}, the limit of strong mixing for $N=1$ yields a distinctly higher Fano factor of $F = 1/3$. We also present numerical simulations for an intermediate regime with $N=3$. Because the $N=3$ results deviate by less than $\sim 3\%$ from the $N \gg 1$ asymptote, we restrict our comparison in the main text solely to the macroscopic limit.

\paragraph{Results: Transmission Eigenvalue Density Distribution.} 
Shot noise probes fluctuations of the transmission eigenvalue distribution, giving more insight than the simple mean conductance. Having a favorable comparison in the main text between the RMT estimates and detailed tight-binding calculations with realistic disorder, we can access more information by examining the whole density distribution $\rho(T)$. Obtaining this distribution directly from tight-binding simulations would require enormous numerical effort; therefore, we use the effective RMT model here as an insightful illustration.

\begin{figure}[ht]
    \centering
    \includegraphics[width=0.8\columnwidth]{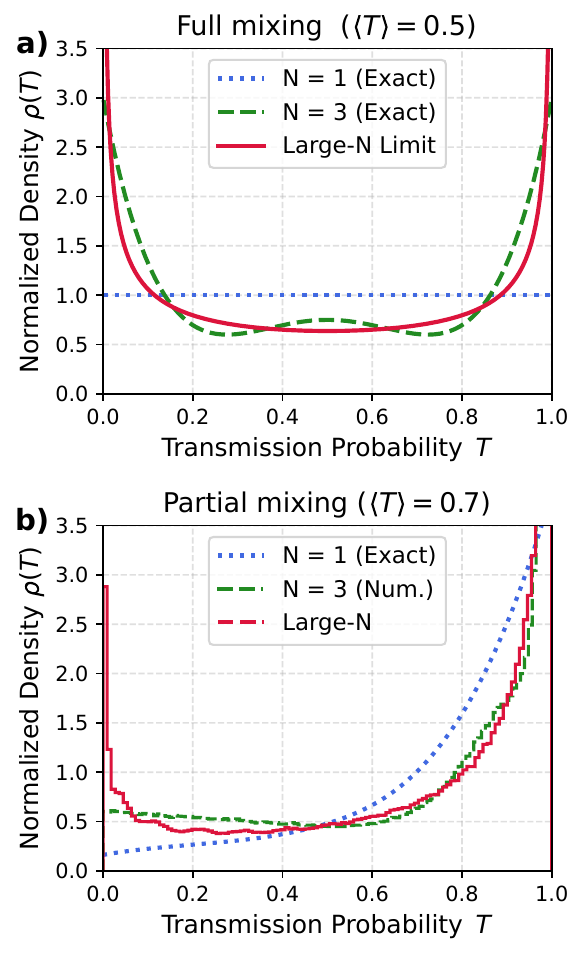}
    \caption{Normalized density distribution $\rho(T)$ of individual transmission eigenvalues. (a) The limit of full mode mixing ($\Gamma \to 0$, corresponding to $\langle T \rangle/N = 0.5$). The distribution is shown for the exact analytical single-channel case ($N=1$), the exact polynomial expansion for $N=3$, and the bimodal macroscopic limit ($N \gg 1$). (b) The regime of partial mode mixing with $\langle T \rangle/N = 0.7$. Solid lines denote the exact analytical result for $N=1$ and the numerical simulation for $N=3$, while the dashed line represents the macroscopic limit.}
    \label{fig:rho_T}
\end{figure}

The limiting case of full mixing ($\Gamma \to 0$) has been worked out in the literature \cite{Brouwer1994, Brouwer1996}. The canonical result for $N=1$ is a flat distribution, which trivially integrates to $F=1/3$. Low mode number $N$ cases can be explicitly expressed in terms of orthogonal polynomials expansion for $\rho(T)$, and then meticulously integrated providing a definite value of $F$. We use the formulas from \cite{Novaes2008} to plot the corresponding polynomial expression for $N=3$ in Fig.\ \ref{fig:rho_T}(a). Integrating this yields $F = 3/95 \approx 0.257$ for $N=3$. Conversely, the limiting case $N \gg 1$ is well known and gives a bimodal density distribution strongly peaked both at $T \to 0$ and $T \to 1$. For comparison and reference, we plot these three results together in Fig.\ \ref{fig:rho_T}(a).

It is instructive to ask how the distribution $\rho(T)$ evolves for a system at some partial mixing. We look in more detail at $\langle T \rangle / N = 0.7$, as we also have data from the microscopic calculations for this value. The distribution for $N=3$ and $N \gg 1$ is not available in a closed analytical form, so we probe it numerically with many samples of random unitary matrix realizations and with some constant coupling $\Gamma$. The result for $N=1$ was calculated analytically in \cite{Brouwer1994}, allowing us to directly plot the exact formulas. These results are collected in Fig.\ \ref{fig:rho_T}(b). For $N \gg 1$, there is some remaining increase in low $T$ values, preserving the bimodal nature of the distribution. Strikingly, the $N=3$ case misses the low $T$ peak, showing barely a slight increase in the $T \to 0$ limit. This is remarkable given that its shot noise is within $3\%$ of the $N \gg 1$ value. Finally, the distribution for $N=1$ exhibits a clear peak at $T \to 1$ but then steadily drops for lower $T$ values; the bimodal characteristic is altogether absent. The shot noise is distinct for $N=1$, as we already discussed and showed in Fig.\ \ref{fig:fano_vs_transmission}.

\paragraph{Conclusion.} 
The relationship between $F$ and $\langle T \rangle/N$ exhibits a robust universality within our simplified model. As demonstrated in the main text, this analytical prediction compares favorably with rigorous numerical simulations incorporating realistic disorder. Consequently, the characteristic $F(T/N)$ dependence proposed here can serve as a quantitative benchmark in future experiments to test the extent to which elastic scattering processes govern shot noise.
\vspace{10pt}
\section*{Acknowledgement}
M.K. acknowledge funding from the Research Council of Finland projects 312295 and 352926 (CoE, Quantum Technology Finland) as well as grant 338872 (NAI-CoG). Our work was also supported by the Ministry of Education of Finland via the Finnish Indian Consortia for Research and Education (FICORE). Our work is part of the QuantERA II Program that has received funding from the European Union’s Horizon 2020 Research and Innovation Programme under Grant Agreement Nos 731473 and 101017733. J.T. received funding from the National Science Centre, Poland, within
the QuantERA II Programme under Grant Agreement Number 101017733, Project Registration Number 2021/03/Y/ST3/00191. We are grateful for fruitful discussions with Ankur Das. We thank the Institute Q and the AScI visiting grant, from which the collaboration originated. The experimental work was performed in Aalto University OtaNano/LTL infrastructures.
\vspace{10pt}
\section*{Author contributions}
J.S designed the sample and performed the experiments, including data collection. S.V, J.S., and M.K. performed the data analysis. S.V. and J. T. performed the theoretical calculations. The results and their interpretation were discussed with S.V., J.S., K.F., J. T., and M. K. All authors participated in writing the article. M.K. supervised the project.

\section*{Data availability}
The tight-binding simulations were performed using 
the open-source \textsc{kwant} package~\cite{Groth2014}, available at \url{https://kwant-project.org}. 
The numerical codes and data generated in this study 
are openly available in the Zenodo repository https://doi.org/10.5281/zenodo.19136560.

\begin{center}
\section*{References}
\end{center}
\bibliographystyle{apsrev4-2}
\bibliography{References}

\end{document}